\documentclass[aps,prl,longbibliography,amsmath,amssymb,twocolumn,superscriptaddress,noeprint]{revtex4-2}
\usepackage{amsfonts,graphicx,bm}
\usepackage{float}
\usepackage{amstext}
\usepackage{amsxtra}
\usepackage{tikz}
\usepackage{footnote}
\usepackage[colorlinks=true,citecolor=blue]{hyperref}
\usepackage{color}
\newcommand{\de}{\partial}

\newcommand{\be}{\begin{equation}}
\newcommand{\ee}{\end{equation}}
\newcommand{\ba}{\begin{eqnarray}}
\newcommand{\ea}{\end{eqnarray}}

\begin{document}
\title{Glitches in rotating supersolids}

\author{Elena Poli$^1$, Thomas Bland$^1$, Samuel J. M. White$^{2,1}$, Manfred J. Mark$^{1,2}$, Francesca Ferlaino$^{1,2,*}$
\\$^1$\textit{Universit\"{a}t Innsbruck, Fakult\"{a}t f\"{u}r Mathematik, Informatik und Physik, Institut f\"{u}r Experimentalphysik, 6020 Innsbruck, Austria} \\$^2$\textit{Institut f\"{u}r Quantenoptik und Quanteninformation, \"Osterreichische Akademie der Wissenschaften, 6020 Innsbruck, Austria}  \vspace{0.2cm}
\\ Silvia Trabucco$^{3,4}$ and Massimo Mannarelli$^{3}$
\\$^3$\textit{INFN, Laboratori Nazionali del Gran Sasso, 67100 Assergi (AQ), Italy} \\ $^4$\textit{Gran Sasso Science Institute, 67100 L’Aquila, Italy}}

\begin{abstract}
Glitches, spin-up events in neutron stars, are of prime interest as they reveal properties of nuclear matter at subnuclear densities. We numerically investigate the glitch mechanism due to vortex unpinning using analogies between neutron stars and dipolar supersolids. We explore the vortex and crystal dynamics during a glitch and its dependence on the supersolid quality, providing a tool to study glitches from different radial depths of a neutron star. Benchmarking our theory against neutron star observations, our work will open a new avenue for the quantum simulation of stellar objects from Earth.
\end{abstract}

\maketitle

One of the greatest strengths of ultracold gases is their ability to simulate the behavior of widely disparate systems \cite{PRXQuantum.2.017003_2}. This extraordinary capability enables quantum gases to serve as powerful solvers for unmasking fundamental open questions concerning the underlying dynamics of complex physical systems. The range of fields where quantum gas simulators have found applications include metallic superconductivity, condensed matter systems, as well as nuclear matter. 
Among these examples, nuclear matter under the extreme conditions existing in neutron stars is the most elusive to direct microscopic observation\,\cite{Shapiro:1983du, Glendenning:1997wn, Haensel:2007yy}.

Neutron stars are the densest stellar objects known today. They form through the core collapse of massive progenitor stars in supernovae type II events, leading to their extreme densities in which a giant gravitational mass of a few solar masses is concentrated in just a tiny radius of about $10$ km. 
Shortly after their birth, neutron stars cool down to temperatures of the order of keV. Compared to ultracold gases (peV), these temperatures are very high, yet much smaller than the MeV energy scale typical of nuclear matter. For this reason, neutron stars can be viewed as cold dense nuclear matter in which quantum effects become very important. The current most-widely-accredited descriptions to explain observations in such systems account for fermionic pairing and correlations in quantum many-body systems\,\cite{MIGDAL1959655, Haskell:2017lkl}.

The 1967 discovery of pulsars \cite{hewish1979observation}--highly magnetized and rapidly rotating neutron stars \cite{gold1968rotating,manchester2005australia}--provided crucial hints of superfluidity and fermionic pairing in these stellar objects. Pulsars can be seen as nearly perfect clocks or regular radio emitters\,\cite{Hamil:2015hqa, Kaspi:2016jkv, Zhou:2022cyp}. They emit photons in a narrow angular beam, similar to that from a lighthouse. This lighthouse effect results from the misalignment between the rotation and magnetization axes and leads to a secular loss of rotational energy with a corresponding slow decrease of the pulsar rotation frequency, $\Omega$. Remarkably, it has been observed that the rotation frequency of the pulsars occasionally shows anomalous jumps--called ``glitches"--in the form of an abrupt speed-up of the pulsar rotation followed by a slow relaxation close to its original value. It is precisely the observations of such pulsar glitches that have provided the first evidence of superfluidity in neutron-star interiors.

This surprising observation suggests  that the interiors of neutron stars are indeed made up of several components, and that one among them is irrotational or at least weakly coupled to the rigid rotation of pulsars. Natural candidates are superfluids and supersolids, respectively. In this scenario, quantized vortices, forming in the superfluid component, can stochastically unpin from the rigid crystalline component and change the star's angular momentum.
Understanding whether this is a plausible mechanism requires addressing several key questions, including: how do superfluid vortices pin and unpin? How do unpinned vortices percolate through the crystalline structure? 
What information can be extracted from the glitch signal shape?

Tackling these questions from first principles is challenging, as the properties of the inner crust of neutron stars are model dependent. Moreover, we only have observational access to the neutron star atmosphere, thus the underlying dynamics are basically a black box. One possible way to improve our understanding of pulsar glitches is to reproduce them in a controllable laboratory, where we have full access to the entire system \cite{tsakadze1973measurement,tsakadze1980properties,graber2017neutron}. 

Thanks to rapid developments in quantum simulation, it is now possible to employ dipolar quantum gases--where supersolidity and rotational physics have recently been observed in circularly symmetric systems \cite{norcia2021tds,Bland2022tds,norcia2022cao,klaus2022oov}--as analogous microscopic quantum systems. Here, we demonstrate exactly this and predict the existence of glitches in a rotating ultracold dipolar supersolid. We show how quantized vortices unpin from the crystalline structure of the supersolid and escape, transferring angular momentum. Varying the interactions, we observe that the glitch size may not only depend on the number of unpinned vortices but also on the superfluid fraction and the supersolid internal dynamics.

\begin{figure}[t]
\includegraphics[width = 0.95\columnwidth]{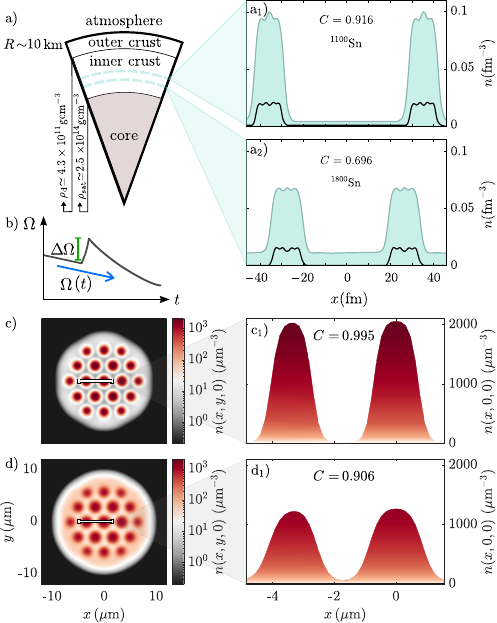}
\caption{Comparison between a neutron star and a dipolar supersolid. (a) Structure of a neutron star, together with the density distributions of neutrons (cyan) and protons (black) near the inner-to-outer crust, for baryonic density $n_b\simeq 5.77 \times 10 ^{-3}$ fm$^{-3}$ (a$_1$) and the inner crust-to-core interface, for $n_b\simeq 2.08 \times 10^{-2}$ fm$^{-3} $ (a$_2$) (adapted with permission from Elsevier from\,\cite{Negele:1971vb}). 
(b) Illustration of a glitch; see text. (c-d) Density distribution of a dipolar quantum gas, with the corresponding density, $n$, cut along $y=z=0$ at (c) $a_s=88 a_0$ and (d) $a_s=93 a_0$, where $a_0$ is the Bohr radius. In both cases, the strength of the superfluid connection is quantified by the density contrast $C = (n_{\rm max} - n_{\rm min})/(n_{\rm max} + n_{\rm min})$.}
\label{fig:comparison_1}
\end{figure}

We start by outlining some basic properties of neutron stars, and then we move to show the analogies with dipolar supersolids. Neutron stars are expected to possess a complex internal structure with a sequence of layers\,\cite{Haensel:1993zw, Lattimer:2000nx, Douchin:2001sv, Haensel:2007yy, Potekhin:2013qqa, Sharma:2015bna,Blaschke:2018mqw, FiorellaBurgio:2018dga}, as shown in Fig.\,\ref{fig:comparison_1}(a).
Beneath a micrometer-thick atmosphere, the first layer, the so-called outer crust, is expected to be a crystalline solid of neutron-rich ions and electrons that behave as a normal component. At its heart, the core of the neutron star is instead believed to be in a liquid-like phase with superfluid properties \cite{Negele:1971vb, PhysRevC.72.015802,  PhysRevC.76.024312, PhysRevC.79.055801, PhysRevC.84.065801, PhysRevC.85.065803}. Here, the density exceeds the nuclear saturation density $\rho_{\text{sat}}$ meaning that the nucleons are so closely packed that they overlap \footnote{The actual composition of the core is unknown: it is believed to be made of  about $90 \%$ of neutrons and $10\%$ of protons and electrons but also muons or other baryons, like $\Delta$ or $\Sigma$, may be present, as well as deconfined quark matter\,\cite{Alford:2007xm, Anglani:2013gfu}.}. 
Sandwiched between the solid outer crust and the superfluid core, one finds the inner crust: here, the density of neutrons exceeds the neutron drip density $\rho_{\text{d}}$ so that it becomes energetically favorable for them to drip out. The most accredited theories describe this phase in terms of unbound superfluid neutron pairs with a periodic density modulation; see Fig.\,\ref{fig:comparison_1}(a$_1$, a$_2$)  and Ref.\,\cite{suppmat}.
The coexistence of solid and superfluid in the inner crust can be viewed in modern terms as a supersolid phase. This, as we shall see, is a key ingredient for the widely accepted physical explanation of glitches, schematically depicted in Fig.\,\ref{fig:comparison_1}(b),  associated with a transfer of angular momentum between the inner and the outer crust\,\cite{BAYM1969n, Ruderman:1972aj, Pines1991, Haskell:2015jra, Zhou:2022cyp,Link:1999ca}.

In the low-energy sector, quantum phases with supersolid properties have recently been observed in various settings \cite{Li2017asp,Leonard2017sfi,Boettcher2019tsp,Tanzi2019ooa,Chomaz2019lla,norcia2021tds,Bland2022tds}. Particularly relevant for drawing analogies with neutron stars is the case of circular supersolids of dipolar atoms \cite{Bland2022tds}, on which we specifically concentrate in this work, as shown in Fig.\,\ref{fig:comparison_1}(c-d).
These systems are obtained by trapping and cooling highly magnetic atoms, like erbium or dysprosium, into quantum degenerate states known as dipolar Bose-Einstein condensates (BECs) \cite{Lu2011sdb,Aikawa2012bec}.
The dipolar supersolid phase exists due to the competition of three types of interactions: a repulsive isotropic contact interaction, a momentum-dependent long-range and anisotropic dipole-dipole interaction and a repulsive higher-order-density interaction arising from quantum fluctuations \cite{chomaz2022dpa}. 
Supersolids are characterized by the existence of a superfluid connection between the crystal sites, controlled in turn by the strength of the short-range interactions, governed by the scattering length $a_s$, which plays the role of the radial depth of the neutron star. Figure \ref{fig:comparison_1}($\text{c}_1$) shows a case with weak superfluid connection, emulating the condition close to the inner-to-outer crust boundary, whereas ($\text{d}_1$) shows one with stronger superfluid connection, in accordance with the inner crust-to-core boundary. 

\begin{figure*}[t]
    \centering
    \includegraphics[width=2\columnwidth]{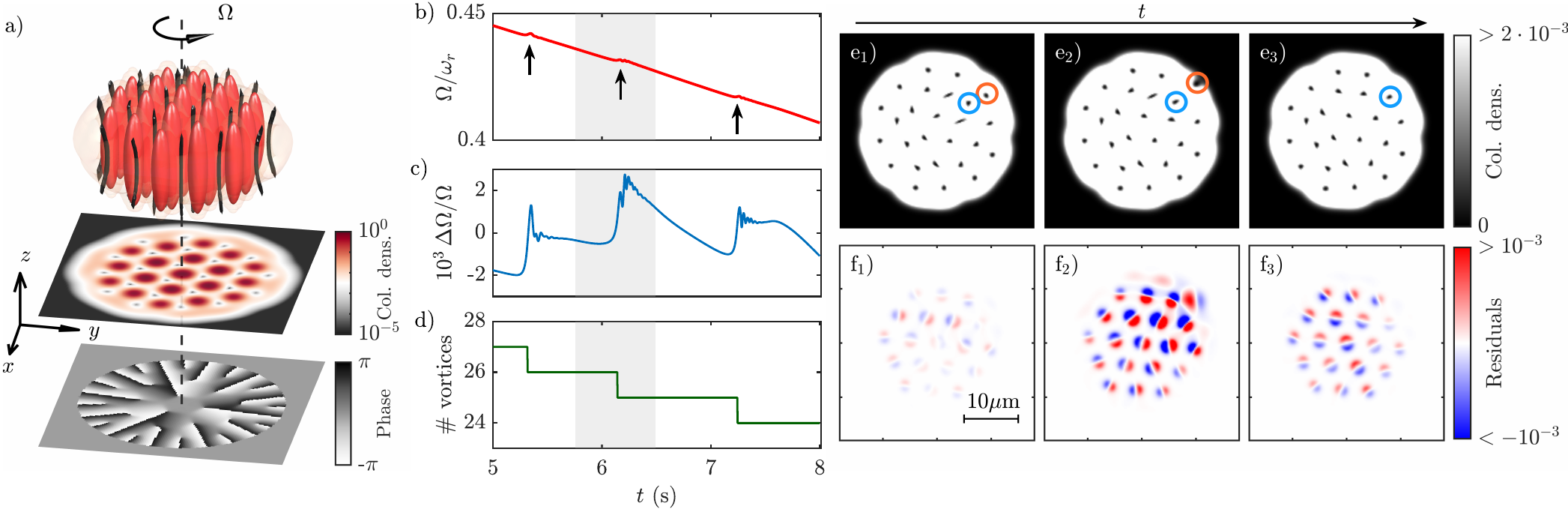} 
    \caption{Glitches in a dipolar supersolid. (a) Rotating supersolid with $\Omega = 0.41\omega_r$ and $a_s = 91a_0$. Top: dipolar supersolid showing two isosurfaces at 15\% (opaque) and 0.05\% (translucent) of the maximum density, and vortex lines in black. Middle: column densities normalized to the peak density. Bottom: phase profile $\arg(\Psi(x,y,z=0))$. (b) Rotation frequency in time, with torque $N_\text{em} = 4.3 \times 10^{-35}$\,kg\,m$^2$/s$^2$. Arrows indicate glitch positions. (c) Relative change in $\Omega$, computed as $\Delta\Omega = (\Omega(t)-\Omega_{\text{lin}})/\Omega_{\text{lin}}$, where $\Omega_{\text{lin}}$ is the result of a linear fit of the curve in (b). (d) Vortex number.  The gray shaded area in (b)-(d) highlights the time window of (e)-(f). (e) Column density saturated to highlight vortex positions and shape, with one vortex escaping (orange circle) and another taking its place (blue circle). (f) Crystal excitations, showing the column density differences between time steps, $n(t) - n(t-\Delta t)$, with $\Delta t = 2.4\,$ms.}
    \label{fig:glitch}
\end{figure*}

The remarkable analogy between a pulsar and a dipolar supersolid can be also extended to the rotational dynamics. In both cases, the time evolution of the rotation frequency, $\Omega$, can be described as\,\cite{BAYM1969n}
\begin{align} \label{eq:glitchmodel}
I_\text{s}\dot\Omega = -N_\text{em} - \dot{L}_\text{vort} - \dot{I}_\text{s}\Omega\,,
\end{align} 
where $I_s$ is the moment of inertia of the solid part. For a neutron star, changes in $I_\text{s}$ are not directly observable and can be challenging to estimate\,\cite{Haskell:2017lkl,Warszawski:2011vy, Warszawski:2012ns, Warszawski:2012wa}. In dipolar supersolids, we have full access to the system, therefore changes in the moment of inertia due to internal dynamics can be accurately accounted for. The quantity $N_{\text{em}}$ is a spin-down torque that linearly reduces the total angular momentum of the star: this process occurs spontaneously in a pulsar due to the emission of electromagnetic radiation, whereas in a dipolar supersolid it can be controlled by slowly ramping down the rotation frequency of the trap. Finally, $L_{\text{vort}}$ is the angular momentum of the superfluid part.

Despite its simplicity, Eq.\,\eqref{eq:glitchmodel} is able to capture very intriguing dynamics in pulsars. While the crystalline part in the inner and outer crust rigidly co-rotates and promptly respond to the braking torque, the superfluid component in the inner crust lags behind, storing angular momentum in the form of quantized vortices. Such vortices are mainly pinned in the interstitial regions, with a pinning force that depends on the depth of the superfluid nuclear background \,\cite{Anderson:1975,Haskell:2017lkl,Negele:1971vb, PhysRevC.72.015802,  PhysRevC.76.024312, PhysRevC.79.055801, PhysRevC.84.065801, PhysRevC.85.065803,Haensel:2007yy}. However, during the spin-down of the star, some vortices can stochastically unpin and escape from  the inner crust, causing a sudden release of angular momentum.
This is captured by the $L_\text{vort}$ term of Eq.\,\eqref{eq:glitchmodel}, which adds a positive contribution to $\dot{\Omega}$ whenever a vortex leaves. 
A glitch corresponds to a collective unpinning of vortices\,\cite{Melatos_2008, Pizzochero_2011}. The outer crust absorbs the released macroscopic angular momentum and suddenly spins up in a step-like fashion, before relaxing and resuming its spin-down behaviour, see Fig.\,\ref{fig:comparison_1}($\text{b}$). The glitches bring a fractional change of the rotation frequency in the range $\Delta \Omega/\Omega \sim 10^{-12}-10^{-3}$\,\cite{Espinoza:2011pq}.

The question now is whether we can validate the above phenomenological description and observe glitches in a dipolar supersolid. To this end, we numerically study the spin-down of an ultracold polarized dipolar BEC in the supersolid state. The atoms with mass $m$ are harmonically confined in a three-dimensional pancake-shaped trap, with frequencies $\bm{\omega} = (\omega_r,\omega_z) = 2\pi\times(50,130)$\,Hz. They interact via the two-body pseudo-potential ${U(\textbf{r}) = (4\pi\hbar^2a_s/m)\delta(\textbf{r}) + (3\hbar^2a_\text{dd}/m)[(1-3\cos^2\theta)/|\textbf{r}|^3]}$, with tunable short-ranged interactions controlled by $a_s$, long-range anisotropic dipole-dipole interactions with effective range given by the dipolar length $a_\text{dd}$, and $\theta$ as the angle between the polarization axis ($z$-axis) and the vector joining two particles. We fix our study to $^{164}$Dy with $a_\text{dd} = 130.8a_0$. The evolution of the macroscopic wavefunction $\Psi(\mathbf{r}, t)$ is governed by the dissipative extended Gross-Pitaevskii equation (eGPE) \cite{Waechtler2016qfi,Bisset2016gsp,FerrierBarbut2016ooq,Chomaz2016qfd}
\begin{align}
i \hbar \frac{\partial \Psi}{\partial t} = (1-i\gamma)\left[\mathcal{L}[\Psi;a_s,a_\text{dd},\bm{\omega}] - \Omega(t) \hat{L}_z\right]\Psi\,,
\label{eq:eGPE}
\end{align}
where $\mathcal{L}$ is the eGPE operator, and we include dissipation through the small parameter $\gamma = 0.05$ to tune the coupling between the system and the rotating trap, see Ref.\,\cite{suppmat}. The wavefunction is normalized to the total atom number through $N = \int \text{d}^3\textbf{r}|\Psi|^2 = 3\times10^5$. The operator $\hat{L}_z  = x\hat{p}_y – y\hat{p}_x$ corresponds to rotation about the $z$-axis, and can be used to obtain the total angular momentum $L_\text{tot} = \langle \hat{L}_z\rangle$. The superfluid angular momentum is obtained from $L_\text{vort} = L_\text{tot} - L_\text{s}$, with the second term $L_\text{s}$ coming from rigid body rotation of the supersolid \cite{Roccuzzo2020ras,Gallemi2020qvi} (see Ref.\,\cite{suppmat}). The initial condition is found in imaginary time, at fixed $\Omega(0) = 0.5\omega_r$, giving a vortex lattice embedded within the supersolid crystal. It has been shown \cite{Roccuzzo2020ras,Gallemi2020qvi,Ancilotto2021vpi} that rotating supersolids host quantized vortices pinned at local minima of the supersolid density modulation, as shown in Fig.\,\ref{fig:glitch}(a), and at saddle points between each pair of droplets \cite{suppmat}.

The real-time spin-down of the system is obtained by simultaneously solving Eqs.\,\eqref{eq:glitchmodel} and \eqref{eq:eGPE}. After generating the initial conditions, we introduce an external torque. This acts as a brake on the solid component, reducing $\Omega(t)$ over time. Our findings are shown in Fig.\,\ref{fig:glitch}(b), where we selected an appropriate time interval to show multiple glitch events. Though at first glance the curve appears linear, dominated by $N_\text{em}$, there are deviations from this behavior highlighted by arrows, showing the appearance of glitches in a dipolar supersolid. Visualizing instead the relative change of $\Omega$ in Fig.\,\ref{fig:glitch}(c), we see signatures similar to pulsar glitches, with a rapid increase of $\Omega$, followed by a slow relaxation back to linear behavior.

Unlike in pulsars, here we have unprecedented access to the internal dynamics of the dipolar supersolid. Thus, we can identify each glitch as the moment when superfluid vortices unpin and reach the trap boundary [Fig.\,\ref{fig:glitch}(d-e)], transferring their angular momenta to the solid component by the feedback mechanism through Eq.~\eqref{eq:glitchmodel}. Furthermore, by tracking the unpinning and re-pinning of individual vortices, we are able to determine the origin of the glitch pulse shape. Here, the observed asymmetry is due to the fact that when internal vortices are unpinned (glitch rise time), it takes some time before they re-pin (glitch fall time): they slowly move from one pinning site to the other, see Fig.\,\ref{fig:glitch}(e$_1$)-(e$_3$)~\cite{suppmat}. Since vortex energy minima are separated by saddle points, to go from one pinning site to the other, a vortex must move across one of them \cite{Ancilotto2021vpi}. In doing this, the vortex core is squeezed and then uncompressed, producing an effective friction on the movement of the vortex. Thus, the long supersolid post-glitch timescale is associated with this slow percolation of vortices across the crystalline structure \cite{suppmat}. As far as we know, this process has never been considered in the description of the pulsar post-glitch behavior. 

We also have access to crystal dynamics. As a consequence of the vortex activity, the crystalline structure is deformed and excited. This is visible in the residual matter density evolution [Fig.\,\ref{fig:glitch}(f$_1$)-(f$_3$)], where, during the glitch, each droplet is slightly deformed and vibrates. Then, during the post-glitch, the droplets slowly relax towards a more uniform distribution. These excitations are due to superfluid fluxes inside the droplets and between neighboring droplets by means of the superfluid bath. Typically, we find that strong crystal excitations affect the post-glitch signal of $\Omega$, suggesting that we could infer the crystal properties through analysis of the glitch pulse shape.

\begin{figure}
    \centering
    \includegraphics[width=0.95\columnwidth]{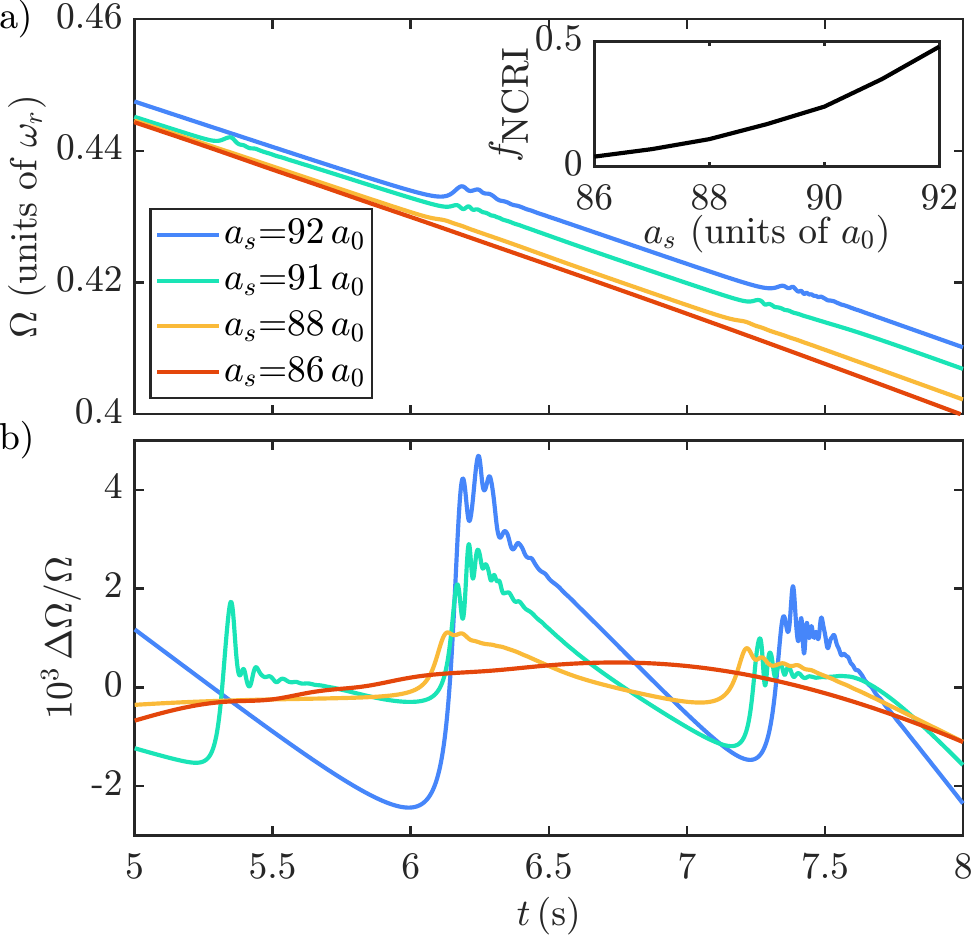} 
    \caption{Glitches originating from different radial depths. (a) Glitches as a function of the scattering length $a_s$. Note, $a_s=92a_0$ emulates the conditions close to the inner crust-to-core boundary, and $a_s=86a_0$ for those in the outer crust. Inset: fraction of non-classical rotational inertia. (b) Relative change in $\Omega$, decreasing amplitude with scattering length. Some glitches dispel more than one vortex, increasing the amplitude.}
    \label{fig:scat}
\end{figure}

The typical magnitude of a glitch is $\Delta \Omega/\Omega \sim 10^{-3}$, a giant glitch in the context of pulsars. The glitch jumps can be written as $\Delta \Omega/\Omega \simeq -\Delta L_\text{vort}/ L_\text{vort}$, as they are dominated by the dispelling of vortices. One may naively expect to estimate $\Delta L_\text{vort}$ as the number of vortices that unpin and reach the boundary multiplied by a quantum of angular momentum $\hbar$. Such an estimate is incorrect because the angular momentum contribution from a vortex is reduced by the fraction of non-classical moment of inertia $f_\text{NCRI}$ \cite{suppmat}, such that $L_\text{vort}$ is at most $f_\text{NCRI}\hbar N_v\le\hbar N_v$ \cite{Gallemi2020qvi}, for the total number of vortices $N_v$. Furthermore, in our finite-size system, the contribution reduces radially from the rotation axis. The combination of these phenomena is such that the effective amount of angular momentum lost by the superfluid component during a glitch is $\Delta L_\text{vort} \simeq 10^{-2} \hbar$ \cite{suppmat}. This suggests that in neutron star glitches, the number of vortices involved in each glitch might be larger than the one estimated by assuming that each vortex carries a quantum of angular momentum.

A reduction of the vortex angular momentum due to the crystal structure also suggests that glitches in the case of vanishing superfluidity will have a small amplitude. We investigate the dependence of the glitch size on the superfluidity by varying the scattering length, as presented in Fig.\,\ref{fig:scat}. As the scattering length is decreased, we find that the glitch amplitude tends to decrease. When the state is in the independent droplet regime ($f_\text{NCRI}\to0$) glitches do not occur. The internal dynamics, though, still slightly affect the response of the system to the external torque, as indicated by the curvature of $\Delta\Omega/\Omega$. The largest glitches occur in the states with the biggest superfluid fraction, and the largest pinning force between droplets. These results suggest that giant glitches in neutron stars occur from deep within the star, where the superfluid contribution to the angular momentum is largest. However, the total amplitude is also reflective of the number of unpinned vortices. The large glitch at 6.2s with $a_s=92a_0$ occurs when two vortices leave together. A possible identifier to discern the origin of the glitches can arise from the post-glitch dynamics, which have the longest decay time at large scattering lengths. 

This work represents a first step in simulating and understanding the complex dynamics of neutron stars using rotating quantum gases in the supersolid phase. We show that these systems exhibit phenomena analogous to neutron star glitches and are primed to become a powerful tool for addressing key open questions ranging from the underlying mechanism of glitches, to the system’s internal dynamics. In particular, during a supersolid glitch, we observe rich dynamics: some vortices unpin and escape towards the outer crust and, in doing so, trigger an excitation of the supersolid crystalline structure, as well as core shape deformation of the remaining migrating vortices. 
These dynamics, which cannot be captured in standard glitch models imposing a fixed lattice structure\,\cite{Warszawski:2011vy, Warszawski:2012ns, Warszawski:2012wa}, could be the key for an experimental implementation of the model, where the dynamical observation of sudden changes in the droplet positions may be possible by combining optimal control methods with non-destructive imaging \cite{van2016optimal,freilich2010real,gajdacz2013ndf}.
Moreover, we see that reducing the superfluidity of the supersolid leads to a reduction of the angular momentum contribution per vortex. This is a feature so far overlooked in the context of neutron stars, and may explain the wide range of observed glitch amplitudes, where the smallest glitches are associated with vortex dynamics at the edge of the star.

Regarding the region of the inner crust close to the core, its investigation requires testing various lattice sizes and vortex configurations, allowing us to expand the study to nuclear vortex pinning expected to occur there~\cite{1991ApJ...373..592L}, akin to the work of Ref.\,\cite{gallemi2022spo}. Furthermore, one could consider systems with a radially variable superfluid fraction to mimic the full structure of the neutron star.
Our work opens the door for a detailed study of the droplet lattice vibration, in order to ascertain whether it is possible to extract the elastic properties of the solid from the supersolid glitch pulse shape. This would be of great astrophysical interest and would pave the way to extract the elastic properties of nuclear matter from the observed neutron star glitch pulse shape, and to test whether a glitch can trigger superfluid collective excitations\,\cite{Andersson_2018}. Finally, future work can investigate the effects of tilting the magnetic field with respect to the rotation axis \cite{prasad2019vortex,klaus2022oov,bland2023vortices}, as expected in pulsars, and include coupling between the supersolid and the proton type-II superconductor present in the crust, through an additional Ginzburg-Landau equation \cite{drummond2017stability,drummond2018stability,thong2023stability}, introducing a self-consistent feedback mechanism. 

We thank Russell Bisset, Wyatt Kirkby, and the Innsbruck dipolar teams for helpful discussions. This study received support from the European Research Council through the Advanced Grant DyMETEr (No.~101054500), the QuantERA grant MAQS by the Austrian Science Fund FWF (No.~I4391-N), the DFG/FWF via Dipolare E2 (No.~I4317-N36) and a joint-project grant from the FWF (No.~I4426). 
E.P.~acknowledges support by the Austrian Science Fund (FWF) within the DK-ALM (No. W1259-N27). T.B.~acknowledges financial support by the ESQ Discovery programme (Erwin Schrödinger Center for Quantum Science \& Technology), hosted by the Austrian Academy of Sciences (ÖAW). S.T. and M.M. thank the Institut für Quantenoptik und Quanteninformation, Innsbruck, for their kind hospitality during the completion of this work. \\
* To whom correspondence should be addressed: \mbox{\url{Francesca.Ferlaino@uibk.ac.at}}.

\clearpage
\appendix
\onecolumngrid
\begin{center}
    {\bf\large Supplemental material: Glitches in rotating supersolids}\\\vspace{0.3cm}
    {\normalsize Elena Poli$^1$, Thomas Bland$^1$, Samuel J. M. White$^{2,1}$, Manfred J. Mark$^{1,2}$ and Francesca Ferlaino$^{1,2}$
\\$^1$\textit{Universit\"{a}t Innsbruck, Fakult\"{a}t f\"{u}r Mathematik, Informatik und Physik, Institut f\"{u}r Experimentalphysik, 6020 Innsbruck, Austria} \\$^2$\textit{Institut f\"{u}r Quantenoptik und Quanteninformation, \"Osterreichische Akademie der Wissenschaften, 6020 Innsbruck, Austria} \vspace{0.2cm}
\\ Silvia Trabucco$^{3,4}$ and Massimo Mannarelli$^{3}$
\\$^3$\textit{INFN, Laboratori Nazionali del Gran Sasso, 67100 Assergi (AQ), Italy} \\ $^4$\textit{Gran Sasso Science Institute, 67100 L’Aquila, Italy}}
\end{center}
\hspace{3cm}
\twocolumngrid
\makeatletter
\renewcommand{\theequation}{S\arabic{equation}}
\renewcommand{\thefigure}{S\arabic{figure}}
\setcounter{equation}{0}
\setcounter{figure}{0}

\normalsize

\renewcommand{\theequation}{S\arabic{equation}}
\renewcommand{\thefigure}{S\arabic{figure}}

\section{Supersolidity in neutron stars}
Nuclear matter in the interior of neutron stars is expected to be in a superfluid state\,\cite{MIGDAL1959655, Chamel:2008ca, Sauls:2019ffv}.
In the outer crust of a neutron star, neutrons and protons form well-defined neutron-rich nuclides. When approaching the interface between the inner crust and the outer crust, the neutron density inside these nuclides increases, the proton population is strongly suppressed and pairing effects between neutrons become sizable. 
The attractive $s$-wave interaction combined with the relatively low temperature may favor the formation of a superfluid state,  resulting in  a lattice comprised of  clusters of superfluid neutrons. 
In the inner crust, corresponding to densities between the neutron drip density $\rho \simeq 4.3 \times 10^{11}$ g cm$^{-3}$ and around the saturation density  $\rho \simeq 2.8 \times 10^{14}$ g cm$^{-3}$, superfluidity leads to interesting effects. Here, nuclear matter consists of connected clumps of approximately one thousand superfluid neutrons and comparatively few protons, surrounded by dripped neutrons forming a ``neutron sea".  At such extreme densities, well-defined nuclides do not exist anymore: the clumps of nuclear matter are sometimes referred to in literature as ``nuclear-type clusters" to emphasize the difference with standard nuclides\,\cite{Baldo:2005nwr}. Nevertheless, it is customary to  associate these clumps with nuclides using their estimated proton number. 
As shown in Fig.\,1($a_1$)-($a_2$) of the main text, the fraction of dripped neutrons  increases with the radial depth (and so, with the density), whilst the overall neutron distribution remains modulated with the periodicity of a crystalline structure. This crystalline structure disappears close to the boundary between the inner crust and the core, where the system becomes homogeneous.

The first calculations of the matter distribution in the inner crust were performed in the Hartree-Fock (HF) approximation, assuming a set of a few non-interacting cells immersed in a sea of neutrons. This distinction was made for numerical reasons and it completely neglected neutron pairing. However, pair correlations play a substantial role in the inner crust\,\cite{Sandulescu:2004xv}. For these reasons, more recent approaches improved the HF calculations using a self-consistent Hartree-Fock-Bogoliubov (HFB) method, combining the HF method with BCS pairing, see for instance \cite{Than:2010tx}. Pair correlations can also be taken into account also using other different approaches, for example the energy functional method developed in Ref.\,\cite{Baldo:2005nwr}.

In the aforementioned works, the density of neutron pairs is found to be modulated within the Wigner-Seitz (WS) cell. 
In fact, pairing effects are smaller in the low-density region (corresponding to the neutron sea) and more relevant in the high-density region (corresponding to the ``nuclides", i.e. the ``solid part").  Thus, in the ``solid part", the pairing and superfluid effects are stronger than in the neutron sea part, because of the higher density. 
In the context of a local density approximation, the pairing field is shown to be a continuous function of position in the whole WS cell\,\cite{Than:2010tx}, supporting the fact that the whole system is superfluid. These results are in agreement with the numerical observation of  excitations in the inner crust, where the appearance of new resonances is due to the collective (``nuclides" + neutron sea) behavior of the system \,\cite{Khan:2004it}. All these works confirm the idea that the ``solid phase" and the ``superfluid phase" are not distinct and, thus, that the system is in a supersolid phase. 

\section{Formalism}
We present here a detailed description of the equation governing the dynamics of an ultracold dipolar Bose-Einstein condensate (BEC) of $^{164}$Dy atoms. Having a large intrinsic magnetic moment,
these atoms interact via a long-range and anisotropic dipole-dipole interaction (DDI).
In the presence of an external magnetic field, the dipolar contribution to the interaction pseudopotential between two atoms reads 
\begin{equation}\label{eq:ddi_simpl}
V_{\mathrm{DDI}}(\textbf{r})=\frac{\mu_0 \mu_m^2}{4 \pi} \frac{1-3 \cos ^2 \theta}{|\textbf{r}|^3}\,,
\end{equation}
where $\mu_m$ is the magnetic moment of the atoms, $\mu_0$ is the vacuum permeability and $\theta$ is the angle between the vector joining the two atoms, $\textbf{r}$, and the polarization axis.
The typical length scale of the DDI is $a_{\mathrm{dd}}=\mu_0 \mu_m^2 m/12 \pi \hbar^2$, where $m$ is the atomic mass and $\hbar$ is the reduced Planck's constant.
In addition to the DDI, magnetic atoms also interact via a short-range contact interaction. This effect is well-approximated by the pseudopotential 
\begin{equation}
    V_{\mathrm{c}}(\mathbf{r})=\frac{4 \pi \hbar^2 a_{s}}{m} \delta(\mathbf{r})\,,
\end{equation}
where the scattering length, $a_s$, is the typical length scale of the contact interatomic force. 
The ratio between the dipolar and contact length scales defines the parameter $\epsilon_{\rm dd} = a_{\rm dd}/a_s$ that allows us to distinguish systems in a contact-dominated regime, where $\epsilon_{\rm dd} <1$, from a dipolar-dominated regime, where $\epsilon_{\rm dd}>1$. In the latter, the system can access the supersolid phase, a paradoxical state of matter that exhibits both superfluid properties and a periodic structure typical of a solid \cite{chomaz2022dpa}. Indeed, as a result of the competition between energy contributions, it is energetically favorable for the system to develop a spontaneous density modulation on top of the superfluid background. For large values of $\epsilon_{\rm dd}$, the superfluid connection between density peaks vanishes, the global phase coherence disappears, and the system enters the independent droplet regime \cite{chomaz2022dpa}.

The ground state and the real-time dynamics of a dipolar gas can be studied by numerically solving the extended Gross-Pitaevskii equation (eGPE):  
\be \label{eq:eGPE}
i \hbar \frac{\de \Psi}{\de t} = (\alpha - i \gamma ) \left[\mathcal{L}[\Psi; a_s, a_{\rm dd}, \boldsymbol{\omega}] - \Omega (t) \hat{L}_z \right] \Psi \,.
\ee
Here, $\Psi(\mathbf{r}, t)$ is the wave function normalized to the total atom number through $N = \int \text{d}^3\textbf{r}|\Psi|^2$ and $\Omega(t)$ is the rotation frequency of the trap about the $z$-axis through the angular momentum operator $\hat{L}_z  = x\hat{p}_y – y\hat{p}_x$. The eGPE operator $\mathcal{L}$ is given by
\begin{align}\label{eq:eGPEoperator}
    \mathcal{L} [\Psi; a_s, a_{\rm dd}, \boldsymbol{\omega}] = & {-\frac{\hbar^2 \nabla^2}{2 m}+\frac{1}{2} m\left[\omega_r^2( x^2+y^2)+\omega_z^2 z^2\right]} \nonumber\\
    & +\int {\rm d}^3 \mathbf{r}^{\prime}\, U\left(\mathbf{r}-\mathbf{r}^{\prime}\right)\left|\Psi\left(\mathbf{r}^{\prime}, t\right)\right|^2 \nonumber\\
    & +\gamma_{\rm QF} |\Psi(\mathbf{r}, t)|^3 - \mu\,,
\end{align}
where $\boldsymbol{\omega}=(\omega _r , \omega_z)=2\pi \times (f_r, f_z)$ are the frequencies of the harmonic confinement with cylindrical symmetry, $U(\mathbf{r})= V_{\mathrm{c}}(\mathbf{r})+V_{\mathrm{DDI}}(\mathbf{r})$ is the total interaction potential, the second to last term is the Lee-Huang-Yang correction ~\cite{Lee:1957zzb}--a beyond mean-field contribution that is particularly important if the system is in the supersolid phase, since it is responsible for its stability against collapse \cite{Chomaz2016qfd,FerrierBarbut2016ooq,Waechtler2016qfi,Bisset2016gsp}--given by \cite{Lima2011qfi,Schuetzhold2006mfe}
\begin{align}
    \gamma_\text{QF} = \frac{128\hbar^2}{3m}\sqrt{\pi a_s^5}\,\text{Re}\left\{ \mathcal{Q}_5(\epsilon_{\rm dd}) \right\}\,,
\end{align}
where $\mathcal{Q}_5(\epsilon_{\rm dd})=\int_0^1 \text{d}u\,(1-\epsilon_{\rm dd}+3u^2\epsilon_{\rm dd})^{5/2}$, and finally, $\mu$ is the chemical potential. When calculating the dipolar potential contribution, we use a spherical cutoff to remove the effect of alias copies coming from the numerical Fourier transform \cite{Ronen2006dbe}.
 
In Eq.\,\eqref{eq:eGPE} the parameters $\alpha$ and $\gamma$ determine the type of evolution:
 \begin{itemize}
     \item $\alpha=0$, $\gamma=1$: imaginary time evolution, to find the ground state of the system.
     \item $\alpha=1$, $\gamma=0$: real-time evolution, to explore the dynamics.
     \item $\alpha=1$, $0<\gamma<1$: complex-time evolution, that corresponds to a real-time evolution with dissipation.
 \end{itemize}
For the purpose of this work, we use imaginary time evolution to generate the initial condition and dissipative real-time evolution for the study of glitch dynamics. In both cases, we employ a split-step method modified to account for rotation, known as the alternate direction implicit-time splitting pseudospectral (ADI-TSSP) method \cite{bao2006aea}, for numerically solving the eGPE. Since the harmonic trap is cylindrically symmetric, the dissipation parameter $\gamma$ is used to impart a rotation to the system, otherwise the angular momentum along the $z$--axis would be conserved during the real-time spin-down evolution.

In addition to the aforementioned method, there are alternative ways to induce rotation in a dipolar supersolid. These include confining the system within an asymmetric trap in the $xy$ plane or utilizing magnetostirring techniques \cite{Martin2017vav,prasad2019vortex,klaus2022oov,bland2023vortices}.

\section{Vortex pinning and dynamics}
Rotating dipolar supersolids host quantized vortices that are pinned in the interstitial low density regions between the droplets. To get a general idea of the pinning energy, we compute the energy cost to imprint a vortex in a specific position ($x_0,y_0$) of the ground state wavefunction $\Psi_0(\textbf{r})$ for a non-rotating supersolid. With this aim, we multiply $\Psi_0(\textbf{r})$ by the ansatz wave function $\Phi_v(x,y,x_0,y_0)$ for a vortex density and phase profile centered at ($x_0,y_0$), given by
\be
    \Phi_v(x,y,x_0,y_0) =  \frac{\sqrt{(x-x_0)^2+(y-y_0)^2}}{\sqrt{(x-x_0)^2+(y-y_0)^2 + \Lambda^{-2}}} e^{i\theta}\,,
    \label{eq:vortexansatz}
\ee
where $\theta = \arctan(y/x)$ and $\Lambda^{-1}=1\,\mu$m \cite{Bradley2012eso}. We note in a dipolar supersolid the phase profile is not an azimuthal 2$\pi$ winding, but rather a complex pattern modified by the underlying crystal structure \cite{Ancilotto2021vpi}, however this simple ansatz recovers the expected force field and stationary points.

\begin{figure}
    \centering
    \includegraphics[width = 0.6\columnwidth]{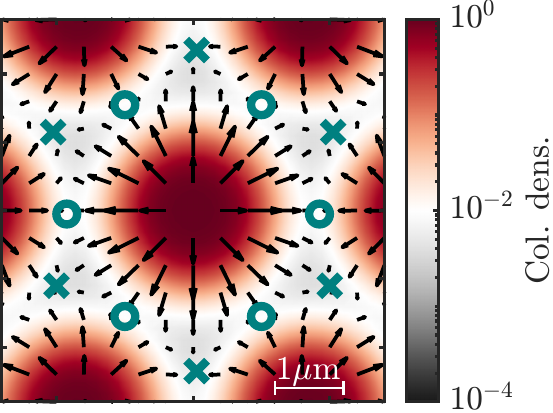}
    \caption{Pinning force felt by a single vortex in a supersolid. The colorbar shows the normalized column density in the central region of a supersolid. The overlaid arrows indicate the direction and the strength of the force imparted to a single vortex in that position. The crosses correspond to the vortex equilibrium positions, and the circles correspond to saddle points. Parameters: $N=3\times10^5$, $a_s=90\,a_0$ and trap frequency $\bm{\omega} = (\omega_r,\omega_z) = 2\pi\times(50,130)$Hz.}
    \label{fig:pinning}
\end{figure}

We compute the total energy $E(x_0,y_0)$ using the wavefunction $\Psi(x_0,y_0) = \Psi_0\Phi_v(x_0,y_0)$ (see, e.g.,~Eq.\,(1) of Ref.\,\cite{Roccuzzo2019sbo}) for different vortex positions ($x_0,y_0$), and the corresponding pinning force $\vec{F}(x_0,y_0) = -\nabla E(x_0,y_0)$. The result is shown in Fig.\,\ref{fig:pinning}. We identify the stable pinning sites -- the absolute minima of the energy landscape -- and the saddle points. Both of them are in the low density regions between the droplets: the former are located in the interstitial sites of the triangular lattice, the latter are between every pair of droplets.

\begin{figure}
    \includegraphics[width=0.85\columnwidth]{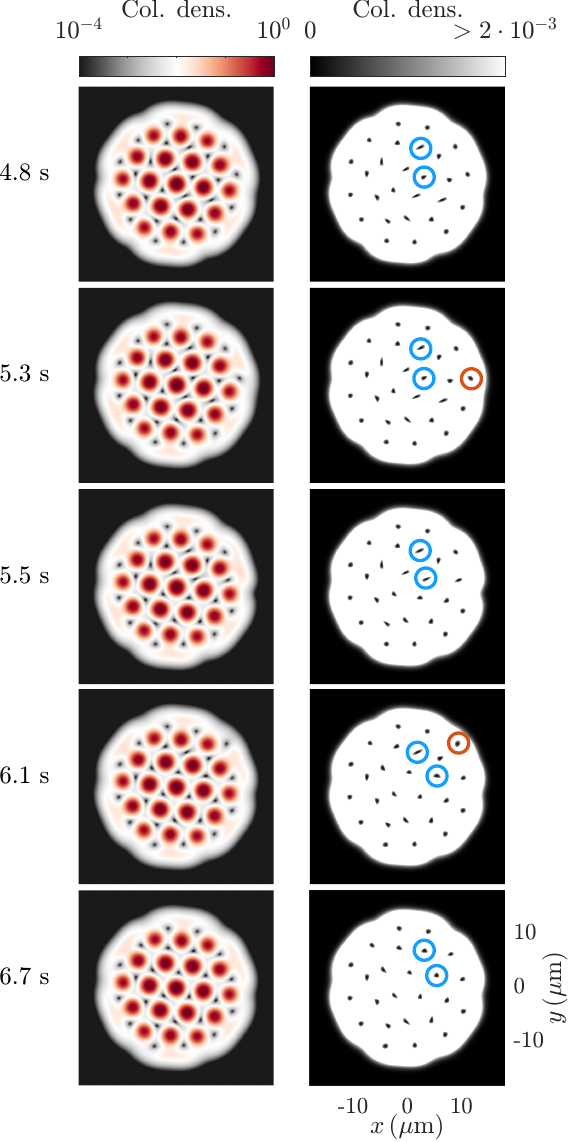}
    \caption{Vortex dynamics during a glitch. Shown are additional frames of the simulation from Fig.\,2 of the main text. Left column: column density normalized to the peak density. Right column: saturated column density, highlighting vortex position and shape before and after glitches. Two vortices (orange circles) escape, while the others (blue circles) rearrange in the lattice.}
    \label{fig:vortex_dynamics}
\end{figure}

This estimate of the energy landscape well approximates the force field acting on vortices during the complex time evolution. The external torque $N_{\rm em}$, see Eq.\,\eqref{eq:LtotNem} and discussion, slowly spins down the dipolar supersolid keeping the position of the droplets almost constant in time in the rotating frame. In the simulation shown in Fig.\,2 in the main text, at around $5.3$ and $6.2$\,s, two vortices escape the system, giving rise to two glitches. Together with these fast dynamics, we observe a slow rearrangement of the other vortices undergoing a force field that resembles the one shown in Fig.\,\ref{fig:pinning}: they can unpin and re-pin from one stable position to another, see Fig.\,\ref{fig:vortex_dynamics}. The blue circled vortices rearrange in space, percolating through the crystalline structure, slowly passing through two different saddle points. As a consequence, vortex-cores appear stretched until they both reach new stable positions.

Finally, Figure \ref{fig:pinning} hints to an unstable maximum at the centre of the droplet. This is expected, as vortices inside droplets are known to be unstable, resulting either in droplet splitting or vortex-line instabilities \cite{PhysRevA.98.023618,PhysRevA.98.063620,PhysRevResearch.3.013283}. Conversely, in neutron stars, vortices can be pinned inside nuclei at sufficiently high densities~\cite{1991ApJ...373..592L}. Furthermore, models considering different scenarios in which the vortices involved are in the core instead of the crust, i.e.~without a solid component, have been discussed in other works \cite{Jones:1998,Mannarelli:2007bs}, but are not considered here.

\section{Feedback mechanism}
\label{sec:angularmom}
We report here the details on the derivation of the feedback mechanism, Eq.\,(1) of the main text. During the dynamics, the constant braking torque $N_\text{em}$ reduces the total angular momentum of the system over time, such that
\begin{equation}
    \dot{L}_\text{tot}(t) = -N_\text{em}\,,
    \label{eq:LtotNem}
\end{equation}
where $L_{\text{tot}}(t) = \langle \hat{L}_z \rangle_{\Psi(t)} $ is the expectation value of the angular momentum operator $\hat{L}_z$ computed for the wave function $\Psi(t)$.
Since the system manifests both solid and superfluid properties, we can decompose the total angular momentum as \cite{Roccuzzo2020ras,Gallemi2020qvi}
\begin{align}
    L_\text{tot}(t) & = L_\text{s}(t) + L_\text{vort}(t) \nonumber\\
                 & = I_\text{s}(t)\Omega(t) + L_\text{vort}(t)\,, \label{eq:Ldecomp}
\end{align}
where $L_\text{s}$ is the angular momentum associated with the crystal rotation, $L_\text{vort}$ is the angular momentum associated to the superfluid and, thus, stored in the form of vortices. The moment of inertia of the supersolid $I_\text{s}$ is time-dependent as well, since the mass distribution of the system changes during the slow-down dynamics.
After inserting Eq.\,\eqref{eq:Ldecomp} into Eq.\,\eqref{eq:LtotNem} and rearranging, we obtain the differential equation for $\Omega(t)$
\begin{align}
    I_\text{s}(t)\dot\Omega(t) = -N_\text{em} - \dot{L}_\text{vort}(t) - \dot{I}_\text{s}(t)\Omega(t)\,,
    \label{eq:glitchmodel}
\end{align}
giving Eq.\,(1) of the main text.

The supersolid moment of inertia is well-defined in the static limit by the definition $I_\text{s,0} = \lim_{\Omega\to0}\langle \hat{L}_z\rangle_{\Psi_{0}}/\Omega$, where $\Psi_0$ is the ground state wave function of the system for vanishingly small values of $\Omega$ \cite{Roccuzzo2020ras,Gallemi2020qvi}. In this limit, it is also pertinent to calculate the rigid body moment of inertia through $I_\text{rigid,0}=\langle x^2+y^2\rangle_{\Psi_{0}}$. The supersolid and rigid moment of inertia coincide if the system is not superfluid and therefore its density distribution fully responds to the external rotation. For a supersolid, this is not the case: the rotational response of the system is reduced because of the superfluid nature, meaning that $I_{\text{s,0}}/I_{\text{rigid,0}}<1$. This lets us define the fraction of non-classical rotational inertia through $f_\text{NCRI}=1-I_\text{s,0}/I_\text{rigid,0}$, which is a quantity closely related to the superfluid fraction \cite{Leggett1970cas}. Therefore, assuming a constant $f_\text{NCRI}$ throughout the simulation, we calculate the time-dependent supersolid moment of inertia through the relation
\begin{equation}
  I_\text{s}(t) = (1-f_\text{NCRI})I_\text{rigid}(t)\,,  
  \label{eq:Is}
\end{equation}
that captures the reduced rotational response of the system and the change in the density distribution at the same time.

\begin{figure}
    \includegraphics[width=0.9\columnwidth]{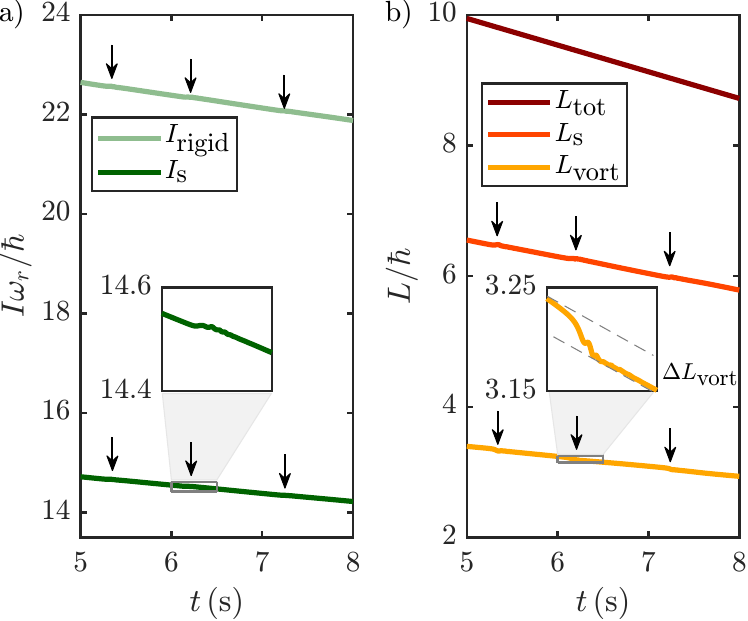}
    \caption{Moment of inertia and angular momenta during the simulation shown in Fig.\,2 of the main text. (a) Rigid moment of inertia $I_{\text{rigid}}$ and supersolid moment of inertia $I_{\text{s}}$. (b) Total angular momentum $L_{\text{tot}}$ and its two contributions from the decomposition $L_{\text{s}}$ and $L_{\text{vort}}$ with arrows pointing to glitches. The inset shows a zoom of the jump $\Delta L_{\text{vort}}$ due to vortices leaving and rearranging during a glitch.}
    \label{fig:Ldec_I}
\end{figure}

In practice, at each time $t$ of the numerical simulation, we compute the total angular momentum $L_{\text{tot}}(t)$ and the rigid moment of inertia $I_\text{rigid}(t) = \langle x^2+y^2\rangle_{\Psi_t}$. As a next step, we compute the time-dependent supersolid moment of inertia $I_\text{s}(t)$ through Eq.\,\eqref{eq:Is}, from which we get the solid contribution to the angular momentum $L_\text{s}(t) = I_\text{s}(t)\Omega(t)$, see Fig.\,\ref{fig:Ldec_I}(a). Then, the vortex contribution to the angular momentum is $L_\text{vort}(t) = L_\text{tot}(t) - L_\text{s}(t)$, see Fig.\,\ref{fig:Ldec_I}(b). Notice that $L_{\text{tot}}$ reduces linearly with gradient $N_{\text{em}}$, as expected. All the necessary quantities are inserted in Eq.\,\eqref{eq:glitchmodel}, providing the updated value of $\Omega$ used as an input to the eGPE shown in Eq.\,\eqref{eq:eGPE}. 
When a vortex reaches the boundary, its contribution to $L_{\rm vort}$ drops to zero due to the negligible matter density around its core, and the linear ramp down of $\Omega $ is interrupted by the glitch event. 
We estimate the glitch size by computing $\Delta \Omega / \Omega =(\Omega  (t) - \Omega _{\rm lin})/ \Omega _{\rm lin} $, i.e. the difference between the observed rotation frequency and the linear fit $ \Omega _{\rm lin}$  that captures the average global spin-down. 

\begin{figure}
      \includegraphics[width=0.95\columnwidth]{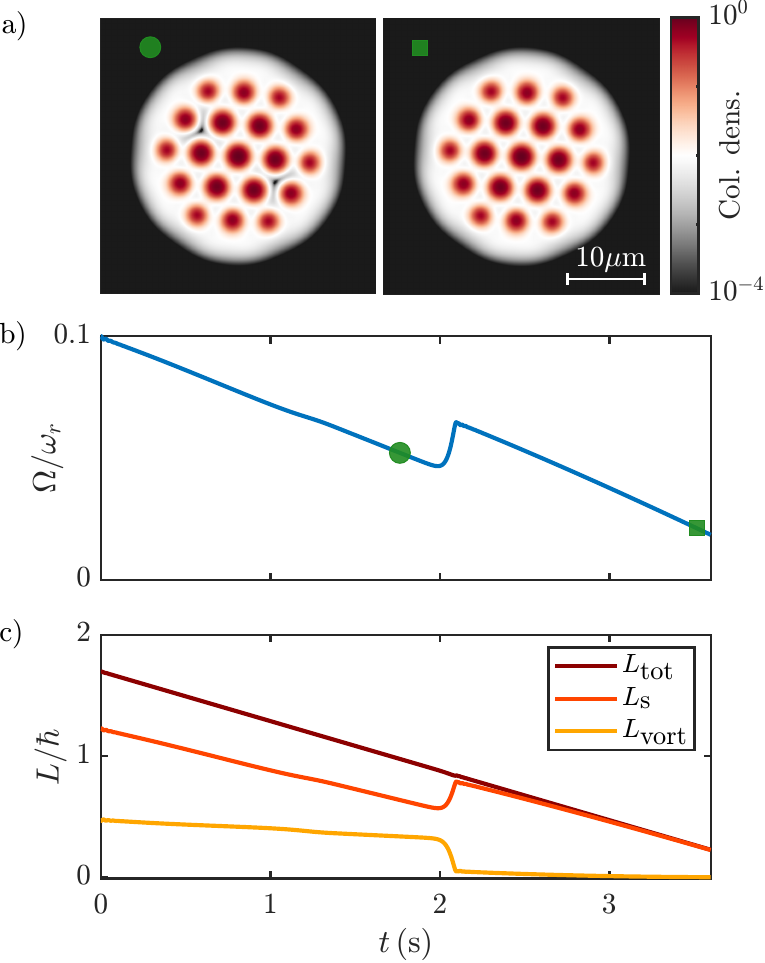}
    \caption{Glitch for low initial rotation frequency $\Omega_0=0.1\omega_r$. (a) Density frames before and after the single glitch. (b) Rotation frequency in time, symbols show the density frames from (a). (c) Angular momentum decomposition: $L_{\rm vort}$ drops to zero when two vortices simultaneously escape.}
    \label{fig:Ldec_01}
\end{figure}

In Fig.\,\ref{fig:Ldec_01} we show a particular case in which around $t\sim2\,$s the last two vortices leave the system resulting in the vortex angular momentum contribution $L_{\text{vort}}$ dropping to 0, thus validating the decomposition [Fig. \ref{fig:Ldec_01}(c)]. It is worth noting also that the crystal structure for $\Omega=0$ shown in Fig.\,\ref{fig:Ldec_01}(a) is unchanged from $\Omega=0.5\omega_r$, validating our decision to fix  $f_\text{NCRI}$.

\section{Robustness of the model}
In the Letter, we have presented simulations for different values of $a_s$, mimicking vortex dynamics for different radial depths in the inner crust of a neutron star. Here, we present additional results at constant $a_s = 91a_0$ with the aim to test the robustness of the model and to identify the appropriate parameter space.

The dynamics of the system for lower rotation frequencies can be studied by setting a smaller value of initial angular velocity $\Omega_0$. The value of $\Omega_0$ primarily influences the number of initial vortices and consequently affects the number of vortices involved in the dynamics during the glitches. For example, only two are present when $\Omega_0 = 0.1\omega_r$ [see Fig.\,\ref{fig:Ldec_01}(a)]. In addition to the results shown in the main text and Fig.\,\ref{fig:Ldec_01}, we performed additional simulations setting the initial conditions to $\Omega_0 = 0.2\omega_r,\,0.3\omega_r,\,0.4\omega_r$ (not shown) observing glitch events analogous to the ones reported and discussed in this work.

\begin{figure}
    \includegraphics[width=0.9\columnwidth]{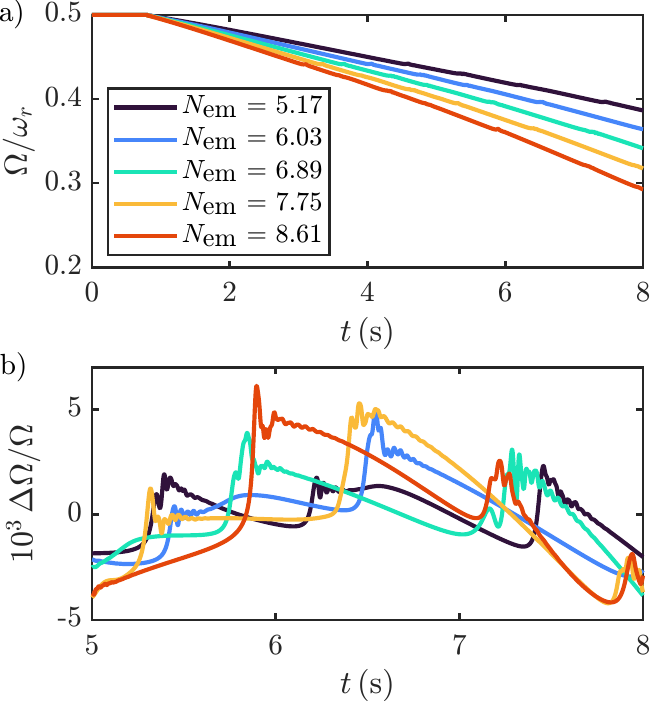}
    \caption{Simulations for different braking torques $N_{\text{em}}$ (units of 10$^{-35}$\,kg\,m$^2/$s$^2$). (a) Rotation frequency in time. (b) Relative change in $\Omega$, computed as $\Delta\Omega = (\Omega(t)-\Omega_{\text{lin}})/\Omega_{\text{lin}}$, where $\Omega_{\text{lin}}$ is the result of a linear fit of the curves in (a).}
    \label{fig:diff_nem}
\end{figure}

We further discuss how the value of the braking torque $N_{\text{em}}$ affects the spin-down dynamics, starting from the same initial condition $\Omega_0=0.5\,\omega_r$ and for constant values of the dissipation parameter $\gamma$.
The results are shown in Fig.~\ref{fig:diff_nem}: 
the glitches occur approximately at the same values of $\Omega \sim 0.44\,\omega_r,0.43\,\omega_r,0.41\,\omega_r,0.39\,\omega_r,0.36\,\omega_r$, albeit reached at different times due to the steepness of the ramp-down process dominated by $N_{\rm em}$. Except for the timescales of glitch events, we do not observe any other significant difference with respect to the simulation showed in Fig.\,2 of the main text.

\begin{figure}
    \includegraphics[width=0.9\columnwidth]{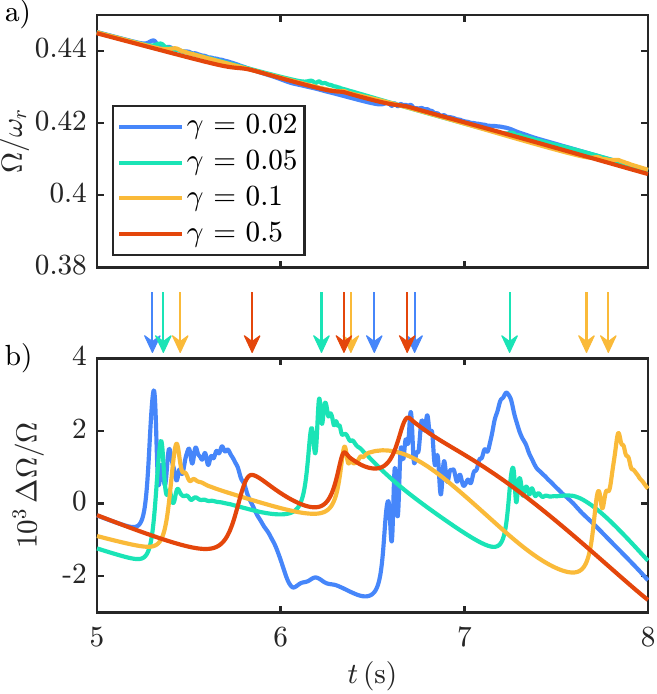}
    \caption{Simulations for different values of $\gamma$.
    (a) Rotation frequency in time, with torque $N_\text{em} = 4.3 \times 10^{-35}$kg m$^2$/s$^2$. (b) Relative change in $\Omega$. The arrows indicate the times in which a vortex is leaving.}
  \label{fig:diff_gamma}
\end{figure}

Finally, we present results for different values of $\gamma$, the parameter responsible for the coupling between the rotating trap and the supersolid. 
The results are shown in Fig.\,\ref{fig:diff_gamma}: we notice that glitches do not always occur at the same time, which is consistent with the fact that the system responds differently to the external torque for varying coupling strengths. Furthermore, $\gamma$ affects the shape of the curve $\Delta\Omega/\Omega$, see Fig.\,\ref{fig:diff_gamma}(b): the oscillations typical of the post-glitch phase are damped ($\gamma=0.1$) or completely absent ($\gamma=0.5$) as the dissipation becomes more relevant.
We notice that for smaller values ($\gamma =0.02$), post-glitch oscillations slow down the recovery process towards a linear ramp down, and the slow rise and fall feature at $t=7.2$\,s is related to an internal rearrangement of vortices, as opposed to a vortex leaving. These features indicate that there exists a threshold value $\gamma \gtrsim 0.02$ such that the eGPE gives a physical description of the rotating supersolid. This is not a surprising fact: we recall that for $\gamma = 0$ the total angular momentum of the system is conserved for any external torque. This is clearly unphysical behavior. Therefore, there must exist a minimum value of the coupling required to have a realistic evolution of the system.

\end{document}